\documentstyle[prb,aps,floats,psfig]{revtex}

\def\be{\beta}

\def\al{\alpha}

\def\lam{\lambda}

\def\Lam{\Lambda}

\def\sof{$SO(5)$ }

\def\D{{\cal D}}
\def\A{{\cal A}}

\def\grad{\vec \nabla}

\def\oDelta{\bar \Delta}
\def\nDelta{ \Delta}

\def\oJ{ \bar J}

\def\vv{{\cal V}}

\def\tc{T$_c$ }

\def\v#1{{\mathbf #1}}

\def\de{\partial}

\def\andword{and }

\def\percent{\%
}

\long\def\taglia#1{#1}

\long\def\tagliasi#1{}

\newcommand{\cmin}{\lesssim}

\def\beq{\begin{equation}}
\def\eeq{\end{equation}}
\def\beqn{\begin{eqnarray}}
\def\eeqn{\end{eqnarray}}

\def\eqref#1{ %
 (\ref{#1})}

\def\andword{and }

\long\def\singlecol#1{
\twocolumn[\hsize\textwidth\columnwidth\hsize\csname @twocolumnfalse\endcsname
              #1]}

\long\def\singlecol#1{#1}

\long\def\singlecol#1{#1}

\def\eqref#1{ Eq.~(\ref{#1})}

\long\def\taglia#1{}

\begin{document}  
\draft   %

\title{
Critical properties of  projected \sof\  models at finite temperatures
}

\author{E. Arrigoni and W. Hanke
}

\address{ 
Institut f\"ur Theoretische Physik,
Universit\"at W\"urzburg,
D-97074 W\"urzburg, Germany \\
e-mail: arrigoni@physik.uni-wuerzburg.de 
}

\singlecol{
 \date{\today} 
\maketitle
\begin{abstract}

We consider the  projected \sof\  bosonic model introduced in
order to connect
 the \sof\  theory of high-T$_c$
superconductivity
with the  physics of the Mott-insulating gap, and
 derive the corresponding effective  functional
 describing low-energy
 degrees of freedom.
At the antiferromagnetic-superconducting  transition,
\sof symmetry-breaking effects due to the gap are purely quantum
mechanical and become irrelevant 
in the neighborhood of a possible finite-temperature multicritical point
separating the normal from the antiferromagnetic 
and the superconducting phases.
A difference in the magnon and hole-pair mobility
 always takes the system away from the \sof-symmetric
fixed point towards a region of instability, 
and the phase transition between 
the normal and the two ordered phases
becomes first order before merging into the
antiferromagnetic-superconducting 
 line.
Quantum fluctuations at intermediate temperatures, while introducing 
symmetry-breaking terms in the case of equal mobilities, tend to
 cancel the symmetry-breaking effects in the case of different mobilities.

\end{abstract}
}%

\section{Introduction}

The  \sof\ theory of high-\tc
superconductivity~\cite{zhan.97,ha.ed.98.fp,ha.ed.98.pc,ed.ha.97}  
has been introduced as a concept to
 unify antiferromagnetism (AF) and d-wave
superconductivity (SC) under a common symmetry principle.
In order to study the physical consequences, 
and to make predictions to compare with experiments,
several exact \sof-symmetric models 
with small symmetry-breaking terms
have been proposed and investigated in
detail~\cite{ra.ko.97,henl.97,sc.zh.97}.
However, a shortcoming of these models, and of an {\it exact} \sof
theory in general  is that they are inconsistent
with the antiferromagnetic gap at half filling, one of the most
important features of the 
high-\tc cuprates \cite{grei.97,ba.an.97u}.  
This can be understood by the fact that an \sof transformation
``rotates'' spin into charge and, thus, a requirement for an
 exactly \sof-invariant system would be to have the same charge and spin
 gap. This is in contradiction with the experimental situation in the
 high-\tc materials, where 
a large charge gap of some eV is present 
in the  AF state 
at half-filling, while spin-wave excitations are ungapped.  
The introduction of a {\it small} symmetry-breaking 
term~\cite{zhan.97,ra.ko.97,henl.97,bu.cl.97,bu.lu.98}, while on the one hand 
correctly selecting the AF state at half-filling and shifting the AF-SC 
transition to finite doping,   does not introduce a charge gap of the 
correct order of magnitude.
In contrast, in a weakly-coupled Hubbard ladder model, a spin and a
charge gap of the same size are present and, in fact, it has been
shown that \sof symmetry is dynamically restored at half
filling~\cite{so5.97,li.ba.98}.  

In order to cure this problem at strong coupling as well,
 a class of \sof\  models -- ``projected'' \sof models --
 has been introduced where the
Mott-Hubbard gap 
is taken into account by means of a 
Gutzwiller projection, whereby doubly-occupied states are projected
out~\cite{zh.hu.99}.
In that , it was shown that,  despite the symmetry-breaking
effects of the projection,
static correlation functions remain exactly \sof\  symmetric within a
mean-field approximation. This is due to the fact that, neglecting
dynamic effects, the Hamiltonian  is 
manifestly \sof\  invariant. However, 
dynamic effects breaking the \sof\  symmetry
become important 
whenever quantum fluctuations are taken into account~\cite{zh.hu.99}.
In another paper~\cite{za.ha.99,ha.za.99.p}, it was 
shown that the projection is crucial in order to correctly
relate the $d$-wave superconducting gap at finite doping with the
$d$-wave modulation of the AF gap observed at half filling by ARPES
experiments~\cite{ro.ki.98}. 

In a microscopic physical system, one would in general expect  \sof\  symmetry
to be explicitly broken by several terms. 
This is certainly the case for the Hubbard model, for example.
However, it  often occurs in nature that a
symmetry, which is broken on the microscopic level, 
is then restored in the long-wavelength limit.
Concerning \sof, this has been shown to happen
for quite generic 
ladder systems of the Hubbard type~\cite{so5.97,li.ba.98}.
Recently, Murakami and Nagaosa
 argued that the bicritical point of the AF to SC 
transition in the  organic superconductor $\kappa$-(BEDT-TTF)$_2$X
(Bis(ethylenedithio)tetrathiafulvalene) with X=Cu[N(CN)$_2$]Cl, 
shows \sof\  critical exponents~\cite{mu.na.99u}.
In fact, one of the  scenarios suggested by Zhang~\cite{zhan.97} is that 
there might be a direct first-order AF to SC transition terminating at a
finite-temperature bicritical point, where the \sof 
symmetry is  asymptotically 
restored at long wavelengths~\cite{zhan.97}.
These ideas are very interesting from an experimental point
of view, and open the possibility of an explicit test of \sof symmetry,
via a direct ``measurement of the number 5''\cite{zhang}.
 This could be done, as suggested in
Ref.~\onlinecite{mu.na.99u},  by measuring the
 critical exponents of the AF to SC transition, which, given 
the spatial dimensionality, should only depend on the number of
components $n$ of
the order parameter.
On the other hand, it is well known that for 
 $n >n_c \approx 4$, the \sof symmetric fixed point is
unstable towards a so-called biconical fixed point~\cite{ko.ne.76}.
 However,
since $n=5$ is close to
 $n_c$, it turns out 
that the stable biconical fixed point only breaks the 
symmetry by about $20\percent$.

However, the situation of the high-\tc materials is quite delicate.
As discussed above,
  the Mott-Hubbard gap plays an important role,
 and  it produces a {\it substantial}
breaking of the \sof\  symmetry.
In Ref.~\onlinecite{zh.hu.99}, it was shown that 
 in the extreme case of a Gutzwiller
projection, a degree of freedom is eliminated completely and the real
and imaginary part of the local superconducting parameter become
conjugate variables.
Therefore, it is not clear whether such a 
{\it projected} \sof symmetry 
can become asymptotically  a {\it complete} \sof symmetry in the
neighborhood of  some critical point.

In this paper, we  show why 
 symmetry-breaking effects due to the projection are asymptotically
irrelevant in the neighborhood of a finite-temperature critical point.
However,
 two kinds of symmetry-breaking effects tend to prevent
\sof symmetry  from being restored asymptotically.
One is related with the 
 different mobilities of hole pairs and magnons ($\eta\not=1$ below), 
and the second one is 
 due to the renormalization effects from quantum fluctuation at an
 intermediate length scale.
The common tendency of these effects is
to draw the system into a region
of instability, where 
the two AF/normal(N) and SC/N transitions become first order before
merging at the AF/SC/N triple point~\cite{ko.ne.76,mu.na.99u}.
However, when the first effect is large, quantum fluctuations tend to
 take the system back to the \sof point.

This paper is organized as follows:
In Sec.~\ref{mode}
we start from the projected \sof-symmetric model (allowing for a
symmetry-breaking term $\eta$ in the mobilities), and treat it by a
slave-boson functional-integral approach in order to deal with the
hard-core constraint.
The important result is that
at the AF-SC transition,
 the {\it classical} part of the action of the
{\it projected} model preserves its \sof structure at $\eta=1$,
despite the symmetry-breaking terms 
arising from the projection.
These terms only appear in the
 {\it quantum-mechanical} (i. e., time
derivative) part of the action.
This fact gives a rigorous justification for the much used
semiclassical description of
the high-\tc materials via a
\sof-symmetric model~\cite{ar.be.97,bu.cl.97,bu.lu.98}  
despite of the
presence of the large Hubbard gap. 
In Sec.~\ref{inte}, we derive the associate
 effective Ginzburg-Landau model by
integrating out the momenta conjugate to the AF superspin
variables~\cite{burg}.
 We study the  properties of such model in the
neighborhood of the AF/SC/N
 triple point and discuss the possibility
of  \sof symmetry restoring at long wavelengths.

In Sec.~\ref{quan}, we evaluate the corrections to the
effective classical action due to the so far neglected quantum
fluctuations. For small temperatures, these mainly affect the
magnon-magnon scattering, thus breaking the \sof symmetry.

Finally, in Sec.~\ref{conc} we draw our conclusions.
Some details of the calculations are given in the appendices.

\section{The model}
\label{mode}

We start from the effective bosonic model introduced in
Refs.~\onlinecite{ed.do.98,zh.hu.99}, 
which describes low-energy {\it bosonic } excitations of  ``blocks'',
(also referred to as ``sites'')
labeled by the coordinate $x$,
 consisting of a rung in a 1-D ladder or of a $2\times2$ plaquette in a 2-D
system.
\begin{eqnarray}
H &=& 
 \oDelta_s \sum_{ x } t_\al^\dagger(x) t_\al(x) + 
 \oDelta_c \sum_{ x } t_i^\dagger(x) t_i(x) \nonumber \\
&-&  \oJ_s \sum_{ <xx'> } n_\al(x) n_\al(x') -
 \oJ_c \sum_{ <xx'> } n_i(x) n_i(x')
\label{anisotropic}
\end{eqnarray}
In this paper, we shall use similar conventions as in
 Ref.~\onlinecite{zh.hu.99}, 
where the indices 
$a,b,..$ are 
the \sof\ superspin indices and 
take the values $1,2,3,4,5$ 
(in some cases, they might also include the ``hole''
index $h$), 
 $\al,\beta,..=2,3,4$ (corresponding to
$x,y,z$) denote the spin indices
and $i,j=1,5$ denote the charge indices, and repeated indices are
implicitly summed over. A boldface sign indicates the vector as a
whole.
Here, $\oDelta_s$ is the energy required to produce a magnon
excitation, i. e. to
replace a singlet
with a triplet in a block, while $\oDelta_c$ is the energy required in order to
produce a particle or hole pair. It is clear that $\oDelta_c$ is of the 
order of the Mott-Hubbard gap and, thus,
$\oDelta_c \gg \oDelta_s$.
  $\oJ_s$ and $\oJ_c$ describe the 
 hybridization 
of these excitations between
nearest-neighbor sites and
are related to their mobility.
The Hamiltonian \eqref{anisotropic} 
acts on a ``vacuum'' $|\Omega\rangle$, which is a kind of
``RVB'' state 
consisting of a product state of 
half-filled singlet states $|\Omega(x)\rangle$ in each block.
On the other hand, the five-fold states $t_a^\dagger(x)|\Omega(x)\rangle$
describe the triplet magnon states (for $a=2,3,4$), and the $d$-wave
hole and 
particle pair states
on a block ($a=1,5$)~\cite{ed.do.98}. 
More specifically, one can define the charge eigenoperators 
$t_h$ and $t_p$ as
\begin{equation}
t_1 = \frac{1}{\sqrt{2}} (t_h + t_p)  \ \ \ 
t_5 = \frac{1}{i\sqrt{2}}(t_h - t_p) \;,
\label{p-h}
\end{equation}
where
 $t_h^\dagger$ is the 
creation operator for a hole pair and $t_p^\dagger$ is the 
creation operator for a particle pair. 
In \eqref{anisotropic}, the $n_a$ play the role of the
``displacement'' coordinates of the local harmonic oscillators, while
we will denote with $p_a$ the conjugate momenta, and we have the
transformation to canonical variables:
\beq
\label{np}
t_{a} = \frac{1}{\sqrt2} \left(n_a + i p_a \right) \;.
\eeq

Due to their microscopic origin these bosonic states 
are hard-core bosons, in the sense that at most one boson can reside
on each site. Mathematically, this is expressed by the condition
\beq
\label{hardcore}
t_{a}^\dag(x) t_{a}(x)  \leq 1 \;.
\eeq 
The particle and hole
 density is controlled by the chemical potential $\mu$, which
couples to the Hamiltonian via a term
\begin{equation}
H_\mu = -2 \mu \sum_x \left[t_p^\dagger(x) t_p(x) - t_h^\dagger(x) t_h(x)\right] \;.
\end{equation}
 In the presence of this
chemical potential term, the gap energy of the hole
and particle pairs become $\oDelta_c + 2\mu$ and $\oDelta_c - 2\mu$
respectively. A (negative) chemical potential of the order of the charge
gap $\oDelta_c/2$ is needed to induce an AF-SC transition
in this system. Near such a transition point, the gap
energy of the hole pair 
 $\oDelta_c + 2\mu$
can be comparable to the (local) spin gap $\oDelta_s$, while the
gap towards a particle pair excitation is pushed up and becomes 
 of the order of twice
the charge gap. Since this is a very large energy scale, we can
safely project this excitation
 out of the spectrum
in the low--energy limit, by requiring that the condition 
\begin{equation}
t_p(x) |\Psi\rangle = 0
\label{constraint1}
\end{equation}
is fulfilled
at every site  $x$. 
The new 
 Hamiltonian takes the form
\begin{eqnarray}
\label{ham}
&&
H =  \oDelta_s \sum_{x,\al} 
    t_{\al}^{\dag}(x) t_{\al}(x) +
   (\oDelta_c + 2 \mu) \sum_{x} 
    t_{h}^{\dag}(x) t_{h}(x) 
\\ \nonumber &&
-
  \oJ_s \sum_{<x,x'>,\al} n_{\al}(x) n_{\al}(x')
\\ \nonumber &&
-
  \oJ_c/2 \sum_{<x,x'>} 
     \left( t_{h}^{\dag}(x) 
    t_{h}(x')  + h.c. \right) \;.
\end{eqnarray}

In Ref.~\onlinecite{zh.hu.99} it was shown that the constraint
\eqref{constraint1} can be enforced by
 introducing canonical commutation rules 
between the two variables $n_1$ and $n_5$, i. e.
\begin{equation}
[n_1,n_5] = i/2 \;,
\end{equation}
and therefore we can identify
$\sqrt2 n_1$ with the ``hole displacement'' $ n_h$ and $\sqrt2 n_5$ with its
conjugate momentum $ p_h$. 
The \sof structure of the Hamiltonian becomes now clear if one
introduces the superspin vector
\beq
\label{ma}
m_a \equiv \left( \eta n_h , n_2, n_3,n_4, \eta p_h \right) \;, 
\eeq
where, for convenience,
 we have absorbed the different mobility for hole pairs and magnons
$\eta\equiv\sqrt{\frac{\oJ_c}{2
     \oJ_s}}$
into the definition of the superspin.
Carrying out the transformation to canonical variables \eqref{np},
the Hamiltonian \eqref{ham} now takes the simple form
\beqn
\label{hso5}
H &=& 
\frac{\nDelta_s}{2} \sum_{ x } p_{\al}(x)^2 
+ \frac{\nDelta_s}{2} \sum_x m_{\al}(x)^2 
+ \frac{ \nDelta_c}{2} \sum_{ x } m_i(x)^2   \nonumber \\
&-&  \ J \sum_{ <xx'> } m_a(x) m_a(x') \;,
\eeqn
where we have further redefined
\beq 
\nDelta_c \equiv \frac{\oDelta_c + 2 \mu}{\eta^2} \quad\quad
\nDelta_s\equiv\oDelta_s \hbox{ , and}  \quad J\equiv \oJ_s \;.
\eeq
The anisotropy in superspin space
 due to $\eta$ reflects now into the constraint, as we will see in
 \eqref{cond} below.
\taglia{
 \eqref{hardcore}, which becomes
\beq
\label{cso5}
 \frac{p_{\al}(x)^2}{2} + \frac{m_{\al}(x)^2}{2} 
+\frac{m_i(x)^2}{2 \eta^2}
\leq 3 \;.
\eeq
}
If one forgets for a moment the connection between coordinates $m_a$ and
their conjugate momenta,
 the Hamiltonian \eqref{hso5} becomes exactly \sof
 invariant under
rotation of the superspin \eqref{ma}
at the AF-SC transition point $\nDelta_s= \nDelta_c$, which is reached by 
changing the chemical potential $\mu$, i. e. at the AF-SC transition.
If one further has $\eta=1$, i. e. $2\oJ_s=\oJ_c$, the constraint is
invariant as well, and one apparently has a complete \sof symmetric model
 (cf. Ref.~\onlinecite{zh.hu.99}).
More specifically, one would like to \sof ``rotate'' just the $m_a$
coordinates, leaving the conjugate coordinates to the magnon part 
$p_{\al}$ unrotated.
This is possible, for example,
 in a 
classical ensemble, where, due to Liouville's theorem,
 expectation values are evaluated 
as $\int \prod_i \ d p_i \ d q_i \exp -H[p_i,q_i] $,
and rotations of the $q_i$ only leaves the measure invariant.
Of course, this 
does not hold for 
dynamics, which is affected by the
relation between
the two
``superconducting'' 
canonically conjugate
components $m_1/\eta$ and $m_5/\eta$, and between 
 the AF components $m_{\al}$ and their conjugate momenta
$p_{\al}$. Thus, 
  \sof symmetry is broken in dynamics,
 as pointed out in
Ref.~\onlinecite{zh.hu.99}. 
Unfortunately, the relation between conjugate variables is also important in
quantum-mechanical {\it static} averages, so that
ground-state or finite-temperature averages are generally 
expected to break the
symmetry when the full quantum problem is taken into account.

In order to understand the nature of the symmetry-breaking terms, it
is convenient to go over to a functional-integral representation of
the partition function for the Hamiltonian \eqref{ham}.
The hard-core constraints can be conveniently taken care of
by means of a
slave-boson representation~\cite{re.ne.85}, where the boson
operator $e(x)$ labeling ``empty'' sites is introduced.
The detailed procedure is shown in Appendix~\ref{slav}.
After this transformation, the action takes the form
\beq
\label{sqmscl}
S=  S_{QM} + S_{CL} \;,
\eeq
where
\beq
\label{sqm}
S_{QM}= \int_{0}^{\beta} d\tau \ \sum_x \left[
 -i p_{\al}(x,\tau) \ \dot m_{\al}(x,\tau)
 -\frac{i}{\eta^2} \ m_5(x,\tau) \ \dot m_1(x,\tau) \right] \;,
\eeq
($\dot m_a$ indicates the time derivative of $m_a$) 
has the well-known  form $p \dot q$ of the Feynman path integral, the
$i$ coming from the 
imaginary-time representation.
Moreover, 
\beqn
\label{scl}
&& S_{CL} = \int_0^{\be} d \tau\ \Bigl\{
\frac{\nDelta_s}{2} \sum_{ x } p_{\al}(x,\tau)^2 
+ \frac{\nDelta_s}{2} \sum_x m_{\al}(x,\tau)^2 
+\frac{ \nDelta_c}{2} \sum_{ x } m_i(x,\tau)^2   \nonumber \\
&-&  \ J \sum_{ <xx'> } e(x,\tau) \ m_a(x,\tau) \ e(x',\tau) \ m_a(x',\tau)
\Bigr\}
\;,
\eeqn
where we have to replace
\beq
\label{e}
e(x,\tau) = \sqrt{1- \frac{p_{\al}(x,\tau)^2}{2} -
  \frac{m_{\al}(x,\tau)^2}{2}  -   \frac{m_i(x,\tau)^2}{2 \eta^2}
}
 \;,
\eeq 
which implicitly includes
the condition 
\beq
\label{cond}
\frac{p_{\al}(x,\tau)^2}{2} +
  \frac{m_{\al}(x,\tau)^2}{2}  +  \frac{m_i(x,\tau)^2}{2 \eta^2}
\leq 1 \;,
\eeq
and where we have already carried out the transformation 
to canonical coordinates,
 \eqref{np}, for the corresponding fields~\cite{discr}.
\eqref{scl}
 is the
 correct classical limit of a {\it projected}, i. e. of the {\it
   physical} \sof model.
Notice that the effects of the hard-core constraint is to
introduce a renormalization of the boson hopping, 
and to bound the superspin magnitude, without, however, fixing
its length~\cite{norm1}.
Thus, the requirement that the superspin magnitude be unity 
should not be taken as a rigorous
 constraint of the \sof theory, at least not of the projected
one (which is the physical one). 
On the other hand,
 one expects that in the
homogeneous ordered phase this constraint  might be a good assumption.
A similar result has been shown by Wegner~\cite{wegn.99u}, namely, that
the orthogonality constraint in the exact \sof model is not a rigorous 
constraint, but it is favored at high temperature, as it maximizes
the entropy. 

\eqref{sqmscl} clearly identifies the \sof-symmetry breaking terms.
The classical action $S_{CL}$ is {\it exactly }
  \sof invariant at the AF-SC transition
($\nDelta_s= \nDelta_c$) and for
$\eta=1$, while apparently incurable
 symmetry-breaking terms come from the
time-derivative terms in
$S_{QM}$.
More specifically, with these values of the parameters, if one 
carries out an \sof rotation within the superspin
vector, \eqref{ma}, $S_{CL}$ remains invariant, while $S_{QM}$ is
changed.
If quantum fluctuations are neglected, 
 one can choose time-independent fields and set \eqref{sqm} to zero.
In this case,
 any equilibrium expectation value
is exactly \sof invariant.
More specifically, let us take a generic \sof rotation matrix
 $R(\v n)= \exp{ i n_a \Gamma_a}$ 
 parametrized by the vector $\v n$ ($\Gamma_a$ are
the \sof generators~\cite{ra.ko.97}), and $f[\v m(x),\v p(x)]$ is
a function of the superspin vector $\v m(x)$, and, possibly, of 
$\v p(x)$. Then, the 
classical
expectation values $< >_{CL}$ have the property
\beq
\label{fmpcl}
< f[\v m(x),\v p(x)] >_{CL} =  < f[\v R \cdot \v m(x),\v p(x)] >_{CL} \;,
\eeq
which is the requirement of \sof invariance. Notice that the $p_{\al}$ 
should not be rotated, while in an exact \sof model they should.

The question is: when is it justified to neglect the time dependence of
the fields ?.
This is allowed 
at moderately high
temperatures, more precisely, at temperatures much larger than
$v/\xi$ (in units of $k_B=\hbar=1$), where $\xi$ is the correlation length
and $v$ is a typical velocity, in our case equal to $J \ a$, $a$ being 
the lattice spacing.
This means that neglecting $S_{QM}$ is {\it exactly justified}
when $\xi$ becomes infinite, i. e. in the neighborhood of
a finite-temperature critical point, as a possible (finite temperature) 
multicritical point
at which the AF/N and the SC/N transition lines 
merge into a first-order line~\cite{zhan.97}.
Moreover, this  critical point is indeed a good candidate for a possible
asymptotic
 restoring of the {\it complete} \sof\ symmetry 
even in the presence, microscopically, of a {\it projected} \sof\ symmetry. 
This is very important as it would mean that the large-energy
symmetry-breaking effect of the
Mott-insulator gap would be exactly
 compensated 
at this critical point.
This  is analogous to the well-known situation
for the antiferromagnetic spin-flop
transition~\cite{zhan.97,ne.ko.74,ko.ne.76,wegn.72,fi.pf.72}
where a system with uniaxial anisotropy restores
$SO(3)$ symmetry at the bicritical point.
However, there are some important differences
with respect to  the spin-flop
transition, as we will show in the next Sections. 
Moreover, notice that due to the symmetry breaking term, \eqref{sqm},
it is unlikely that
\sof symmetry
 can be restored 
if the AF-SC transition is controlled by a quantum-critical point.
Since we are interested in finite-temperature  critical points, we
will restrict to   the case of three spatial dimensions $D$.

\section{Effective Ginzburg-Landau
  action}
\label{inte}

In this Section, we study the
 action \eqref{hso5} in more detail.
We first integrate out the momenta $p_{\al}$ and obtain
an effective action restricted to the superspin variables.
For  temperatures smaller than the singlet-triplet splitting
$\Delta_s$,
 one can restrict to a Gaussian
integration of the momenta, i. e., consider only quadratic terms in
$p_{\al}$.
Carrying out such an expansion, one obtains
\beq
S_{CL} = S_{pm} + S_m + {\cal O}(p_{\al}^4) \;,
\eeq
where, leaving the $\tau$ dependence implicit 
\beqn
\label{sm}
&& S_m = \int_0^{\tau} \ d\ \tau \ \bigl[ 
 \frac{\nDelta_s}{2} \sum_x m_{\al}(x)^2 
+\frac{ \nDelta_c}{2} \sum_{ x } m_i(x)^2   
\nonumber \\
&-& 
 \ J \sum_{ <xx'> } r(x) \ m_a(x) \ r(x') \ m_a(x') 
\bigr]
 \;,
\eeqn
\beq
\label{spm}
 S_{pm} = \int_0^{\tau} \ d \tau \sum_{ x } 
 \ \frac{\nDelta_s}{2} \ \A(x)\ p_{\be}(x)^2 \;,
\eeq
where we have defined
\beq
\label{cala}
\A(x) \equiv
1
 +  \ \frac{2\ J}{ \nDelta_s}  \frac{m_a(x)}{4 r(x)}    
\sum_d^{nn} m_a(x+d) r(x+d)
 \;,
\eeq
\beq
\label{rx}
r(x) \equiv \sqrt{1-\frac{m_{\al}(x)^2}{2}- \frac{m_i(x)^2}{2\ \eta^2}} \;,
\eeq
and the sum $\sum_d^{nn}$ extends over nearest-neighbor sites.

It is now convenient to reabsorb the $\v m$-dependent 
coefficient $\A(x)$ of the $\v p^2$
term
into the definition of the momenta $\v p$. This is done in order to avoid 
the appearance 
of  terms 
depending on
 the amplitude of the imaginary-time slice
in the effective action.
Furthermore,
in order to avoid a $\v m$-dependent Jacobian due to the transformation,
it is  convenient to transform the $\v m$-coordinates
in such a way 
  that the Jacobian remains unity.
The general procedure is illustrated in Appendix~\ref{mome}.
Up to second order in $\v m^2$, the new $\v m'$ coordinates 
are related with the old ones via
\beq
\label{mamap}
 m_a(x) = m_a'(x) ( 1+ \frac{3 J }{7 \nDelta_s} |\v m'(x)|^2 ) \;.
\eeq

After this transformation, the integration of the $p'_{\al}(x)\equiv
p_{\al}(x) \sqrt{\A(x)}$ only affects $S_{QM}$, 
 and one obtains a new QM action in the form
\beq
\label{sqmp}
S_{QM}' = \int_0^{\be} \ d \tau \int d x \ \left[ \frac{ \dot m_{\al}^2 }{ 2
  \nDelta_s  \ \A(x) } - \frac{i}{ \eta^2} \ m_5(x) \ \dot m_1(x) \right]\;,
\eeq
where the transformation \eqref{mamap} should be inserted, and 
we have absorbed the unit cell volume
$\vv=a^3$ in the definition of the fields by renaming $m_a^2/\vv \to m_a^2$.

Thus, the total effective \sof action restricted to the superspin variables
is given by \eqref{sm} plus \eqref{sqmp}.  The transformation \eqref{mamap} 
must still
be carried out on the $m$ variables, but, due to the fact that the
coefficient $\frac{3 J}{7 \Delta_s}$ is small at the transition,
this does not change the result
  significantly. 
On the other hand, it is important to take into account the effects
of the hard-core constraint, which introduces the transformation
\eqref{rx}, and, implicitly, 
the restriction of the superspin within a $5$-dimensional hypersphere
(or a ellipsoid, if $\eta\not=1$) \cite{norm1}.
Thus,
 $S_m$ (\eqref{sm}) gives the 
 first  effective classical functional
microscopically derived from an \sof model~\cite{burg},
 where the physics of the
Mott insulating gap has been properly taken into account via the projection.
This is the appropriate 
functional
which should be used for {\it physical} predictions 
of the \sof theory, consistent with the gap.

Close to the phase transitions, it is more convenient to derive
a Ginzburg-Landau form for the action, 
obtained, as usual, by expanding in powers of the field $\v m$
and keeping only lowest-order gradient terms.
After
 inserting  \eqref{mamap} and
dropping the prime indices in the fields $\v m$, we obtain
\beqn 
&&
\label{sgl}
S_{CL}' =   \int_0^{\be} d \tau \ \int d x\  \Bigl\{ 
\frac{r_s}{2}  m_{\al}(x)^2 +
           \frac{r_c}{2} m_{i}(x)^2 + 
\frac{\rho}{2} (\grad m_a(x))^2 
\\ \nonumber &&
+ \frac{u_s}{8} \left(\sum_{\al} m_{\al}(x)^2\right)^2 +
\frac{u_c}{8} \left(\sum_{i} m_{i}(x)^2\right)^2 +
\frac{u_{cs}}{4} \left(\sum_{\al} m_{\al}(x)^2\right) \
 \left(\sum_{i} m_{i}(x)^2\right) 
\Bigr\} \;,
\eeqn
with the parameters
\beqn
&&
\label{param}
\ \frac{r_{s/c}}{2} =  (\frac{\nDelta_{s/c}}{2} - D  J)
\\ \nonumber &&
\ \frac{\rho}{2} = \frac{J \ a^2}{2} 
 \\ \nonumber &&
\ \frac{u_s}{8} =   \frac{\vv \ J}{2} (D + \frac{3\ r_s}{7\ \Delta_s}) 
 \\ \nonumber &&
\ \frac{u_c}{8} =   \frac{\vv \ J}{2} (\frac{D}{\eta^2} + \frac{3
 \  r_c}{7\ \nDelta_s}) 
 \\ \nonumber &&
   \ u_{cs} = \frac{u_c+u_s}{2} \;.
\eeqn
\taglia{
where we have
considered the fact that~\cite{aniso}
\beq
\vv \sum_x m(x) \sum_d^{nn} m(x+d) = \frac{\vv}2 \sum_{x}\sum_d^{nn}
\left[ m(x)^2 + m(x+d)^2 - \left(m(x)-m(x+d)\right)^2 \right]
\approx 2 D\ \int d x\ m(x)^2 - a^2 \ \int d x\  \
(\grad m(x))^2 \;.
\eeq
}

The critical properties of the model \eqref{sgl} have been analyzed
in several
works~\cite{ko.ne.76,ne.ko.74,fi.pf.72,ke.wa.73,wegn.72,ahar.73,mu.na.99u}.
Its phase diagram
is determined  by two relevant
parameters, the first one $r_s-r_c \propto \Delta_s-\Delta_c$ controls the
transition between the AF and the SC phases, while the other $\sim
\min(r_s,r_c)$ controls the second-order transition between the 
appropriate ordered (AF or SC) and  the normal
phase.
At the transition point
 $r_s \sim r_c\sim 0$, there are two competing fixed point controlling the 
transition~\cite{ko.ne.76}, the Heisenberg bicritical fixed point (in this
specific case,
the \sof fixed point), and the biconical tetracritical fixed point.
According to the
$\epsilon$-expansion,
the latter fixed point turns out to be  the stable one for $n > n_c
\approx 4-O(\epsilon)$.  
This means that, in general,  the model \eqref{sgl}, which has $n=5$,
is expected to flow to this latter fixed point and not to the
 \sof-symmetric one for $u_s\not= u_c \not= u_{cs}$.
On the other hand,
 since $n=5$ is not very far away
from $n_c$, the stable biconical fixed point is
 approximately
 \sof invariant
with symmetry-breaking terms of the order of $20\percent$.
Moreover,
there is a plane in the $u_s,u_c,u_{cs}$ space, given by the condition
$u_{cs}^2 = u_c u_s$,
 from which 
the system  flows to
the \sof point~\cite{ko.ne.76,mu.na.99u}. 
This is due to the fact that a
 scale transformation of, say, the SC components $m_i^2 \to m_i^2
 u_s/u_c$ 
of the order
parameter would yield again 
an \sof-symmetric interaction of the form $u |\v m|^4$.
The asymmetry would then be transfered
into  different 
susceptibilities $\rho_s,\rho_c$ for the AF and for the SC order
parameters.
However, it has been shown  in Refs.~\onlinecite{pe.ne.76,wegn.72}
that the different in the susceptibilities is an 
irrelevant parameter.

In our case, we have
\beq
\label{ucsucs}
\Delta u^2 \equiv u_{cs}^2 - u_c u_s = \left(\frac{u_c-u_s}{2}\right)^2 \geq 0
\eeq
which means that the \sof symmetric fixed point is never reached,
 except
when the equal sign holds, i. e., when $\eta\not=1$ (at the transition 
$r_s=r_c$).
On the other hand,
we expect on physical grounds 
the mobility of the hole
pairs to be
 smaller than that of the magnons, and, thus, $\eta$ to be
 smaller than $1$.
Unfortunately,
for the case \eqref{ucsucs}, the couplings flow away into a
region of instability. The common interpretation  is that 
 the AF/N and SC/N transitions become first order as well
(fluctuation-induced first-order transition), 
at least close
enough to the AF/SC/N triple point~\cite{ko.ne.76,mu.na.99u}.

This fact seems in contrast with the apparent observation of bicritical
behavior with \sof critical exponents 
in  the organic superconductor
$\kappa$-(BEDT-TTF)$_2$X (see Refs. ~\onlinecite{ka.mi.95,ka.mi.97}), 
by Murakami and Nagaosa~\cite{mu.na.99u,dynam}.
There may be several ways to understand this.
One possibility is that other effects  not considered here, such as,
e. g., Coulomb 
interactions, fermionic excitations~\cite{ha.za.99}, 
or quantum effects, as discussed in Sec.~\ref{quan},
 counterbalance this effect and draw the system back to 
 the domain of attraction of the biconical fixed point.
As discussed above, the differences between the \sof and the biconical
fixed point  
are only about $20\percent$, so that they might be not observable
experimentally.
Alternatively, since 
 the flow would cross
the \sof plane, it could produce \sof exponents 
at intermediate
 length scales.
On the other hand,
 Hu et al.~\cite{hu.ko.99,hu.99u}, 
observe a coexistence region of AF and SC
for the \sof-anisotropic case, which could be possibly
identified with the biconical phase. 
Their result could be
due to the fact
that they consider a different $c$-axis
 anisotropy 
 ($\chi$ in
 footnote \onlinecite{aniso}), for the AF and for the SC variables.

\section{Quantum corrections}
\label{quan}

Even when considering a classical (i. e., finite-temperature) critical 
point, the quantum-mechanical symmetry-breaking terms $S_{QM}$
although irrelevant in the RG sense, contribute 
to the RG flow
up to a certain length 
scale of the order  of $v/T$.
Since $S_{QM}$ breaks the \sof symmetry, it is expected, during this
initial renormalization process, to introduce symmetry-breaking terms in
$S_{CL}$.
Therefore, even when $S_{CL}$ is \sof symmetric at the microscopic
scale,
the renormalized $S_{CL}$ at the scale $\xi\sim v/T$ will 
probably break the symmetry.
In this Section, we evaluate these symmetry-breaking terms originating 
from $S_{QM}$, or, more precisely, from 
 the time dependence of the fields.

In order to evaluate these effects, we separate the fields into their
static  and dynamic parts, and integrate out the latter.
Since we are working at finite temperature, we have to integrate out
the components of the fields with Matsubara frequencies $\omega_n=2
\pi n T$ with $n \not =0$. 
In order to obtain an analytic expression for these corrections, we
restrict to one-loop contributions and take just the leading
low-temperature terms.

We first diagonalize the non-interacting (quadratic) part of the action
\eqref{sqmp} plus \eqref{sgl} by Fourier transform. We can neglect
the corrections to $S_{QM}'$ due to the transformation to the primed
variables \eqref{mamap}, as it introduces irrelevant quartic
time-derivative terms.
In Fourier space, the action takes the usual form
\beq
\label{sfourier}
S_{QM}' + S_{CL}' = \frac{1}{2} \ {\sum \hspace{-1.35em} \int}_k 
m_a(-k) \left[G(k)^{-1}\right]_{a b} m_b(k) +
\frac{1}{8} {\sum \hspace{-1.35em} \int}_{k_1,k_2,k_3}
m_a(k_1) \ m_a(k_2) \ u_{a b} \ m_b(k_3) \ m_b(-k_1-k_2,-k_3) \;,
\eeq 
where we have introduced the shorthand notation 
$k \equiv (k,\omega)$, and 
${\sum \hspace{-1.em}  \int}_{k}
 \equiv \frac{1}{\be} \sum_{\omega} \int^{\Lambda} \frac{d^3 k}{(2 \pi)^3}
$, with $\Lambda\sim 1/a$ a short-distance cutoff for $k$. 
In \eqref{sfourier},          
 the  nonzero elements of the (non interacting) Green's functions read
\beq
\label{gab}
G(k)_{\al \be} = \frac{\delta_{\al,\be}}{
r_s+\frac{\omega^2}{\Delta_s} + \rho k^2 }\;,
\eeq
\beq
\label{g11}
G(k)_{1,1}=G(k)_{5,5}= \frac{ \rho k^2 + r_c}{ (\rho k^2+r_c)^2 +
  \frac{\omega^2}{\eta^4}} \;,
\eeq
and
\beq
\label{g51}
G(k)_{5,1}=-G(k)_{1,5}= \frac{ \frac{\omega}{\eta^2}}{ (\rho k^2+r_c)^2
  + 
\frac{\omega^2}{\eta^4}} \;,
\eeq
and the interaction parameters are
 $u_{\al,\be} = u_s$, $u_{i,j} = u_c$,  and $u_{i,\al}=u_{cs}$.

At one loop, 
 integration of the $\omega \not =0$ fields 
only  changes
 the parameters $r$, and $u$, similarly to  conventional field theory.
In the $T\to 0$ limit, the change of the former is finite, and 
merely shifts the transition point.
On the other hand, the changes $\delta u_{ab}$ 
in the interaction parameters $u_{ab}$ grow logarithmically
with decreasing temperature at the critical point.
We will, thus, restrict to evaluation of these corrections.
These are given by the sum of the usual ``loop'' diagrams, which give
\beq
\label{duab}
\delta u_{a b} = -\frac12 \sum_c \ u_{a c} \ u_{c b} \ I_{cc} - 2 \ u_{a b}^2
\ I_{a b}
- u_{a b} \ \left( I_{aa} \ u_{aa} + I_{bb} \ u_{bb} \right) \;,
\eeq
where the integrals $I_{ab}$ are given by
\beq
\label{iab}
I_{a b} = {\sum \hspace{-1.35em}  \int}_{k,\omega\not=0}
   G(k)_{a a} G(-k)_{b b} \;.
\eeq
In \eqref{duab} and \eqref{iab}, we have neglected
contributions from 
 nondiagonal parts of Green's functions~\eqref{g51}, as they only give finite
 contributions to integrals of the form \eqref{iab} 
in the low-temperature limit.
The same holds for integrals containing at least one Green's function
of the superconducting fields~\eqref{g11}. This is due to the fact that for these
fields the (bare) dynamical critical exponent~\cite{ho.ha.77,hert.76}
$z$ is equal to $2$, and it does not produce divergences in $D=3$.
This in turns occurs because the two components of the SC
order parameter are canonically conjugate, while the AF ones have
independent massive ones.
Therefore, we will consider only the divergent contribution 
\beq
\label{ialbe}
I_{\al,\be} = I_s \equiv \frac{1}{8 \pi^2} \sqrt{\frac{\Delta_s}{\rho^3}} \ln
\frac{\Lam \sqrt{\rho \ \Delta_s}}{2 \pi T} \;,
\eeq
where we have assumed that we lie outside of the region of influence
of the quantum critical fixed point, i. e.
$T \gg \sqrt{ r_{c/s}  \Delta_s}$.
Replacing \eqref{ialbe} in \eqref{duab}, we obtain for the leading
contributions 
\beqn
\label{deltau}
&&\delta u_s = - \frac{11}{2} \ u_s^2 \ I_s
\\ \nonumber &&
 \delta u_c = - \frac{3}{2} \ u_{cs}^2 \ I_s
\\ \nonumber &&
\delta u_{cs} = - \frac{5}{2} \ u_{s} \ u_{cs} \ I_s \;.
\eeqn
As expected,
 quantum fluctuations
draw the system away from the \sof-invariant point even in the case
where $\eta=1$. 
This can be seen by adding these corrections to an initially
 \sof-invariant system with $u_c=u_s=u_{cs}=u$.
At the lowest order in the $u_a$,
the renormalized parameters  
$u_a'=u_a+\delta u_a $ obey the relation
\beq
\label{ucsp}
\Delta u^{'2} \equiv u_{cs}^{'2} - u_s' \ u_c' = 2 \ I_s \ u^3  >0 \;,
\eeq
i. e., as in the case of $\eta\not=1$, \eqref{ucsucs},  the system is drawn 
into the instability region where a fluctuation-induced first-order
transition is expected.
This indicates that quantum fluctuations and
anisotropy $\eta\not=1$ cooperate in the same direction and draw the system
into the instability region, where no finite fixed point is expected.
However, for the case where the $u_a$ are different, one obtains 
\beq
\label{ucspd}
\Delta u^{'2} - \Delta u^{2} = 
- \frac{u_s}{2} \left( 7 u_{cs}^2 - 11 \ u_c \ u_s \right) I_s \;.
\eeq
Further 
 inserting the values of the
 $u_a$ from \eqref{param}
with  $\eta\not=1$
 (we fix ourselves at the triple point $r_s=r_c$),
\eqref{ucspd}
 becomes negative for $\eta<x_c$, or $\eta>1/x_c$
with $x_c\approx 0.498$. 
Therefore, for large difference in the mobilities $\eta$, quantum
fluctuations tend to shift the renormalized parameters back towards
the domain of attraction of the biconical and of the \sof fixed point.

\section{Conclusions}
\label{conc}

In conclusion, we have analyzed the properties of a projected \sof
model which takes into account the high-energy physics of the
Mott-insulating gap.
As already pointed out in Ref.~\onlinecite{zh.hu.99},
the chemical potential can always be shifted to the AF-SC transition point
in order to cancel the symmetry-breaking terms
produced by the gap
in the classical part of the action.
On the other hand,
symmetry-breaking terms due to the projection 
show up in the quantum-mechanical 
part of the action, as a
 conjugacy relation 
between the superconducting components of the superspin vector. 
A further
source of symmetry breaking is due to the different
mobility of the hole pairs and of the magnons parametrized by $\eta\not=1$.

Close to the AF/SC/N finite-temperature multicritical point, the quantum
effects due to the projection are irrelevant, although subleading
symmetry-breaking corrections appear at intermediate length scales.
When considered separately, these symmetry-breaking effects both draw
the RG flow into a region of instability 
with first order transitions and
no \sof symmetry.
On the other hand,
for strong anisotropies
$\eta\cmin 0.5$,  quantum 
corrections
partly cancel
the symmetry-breaking effects. 

There are possibly other effects, such as Coulomb
interaction, or fermionic excitations, 
which can possibly take the system back into the domain of 
attraction of the biconical fixed point, where \sof symmetry is only
broken by $\sim 20\percent$.
Notice that, since the order parameter
must be rescaled in order to reach this fixed point, the (possibly
approximate) \sof symmetry 
reached at this critical point is {\it renormalized}, in the sense of
Ref.~\onlinecite{so5.97}.
This means, for example, that the \sof picture would be consistent with
different absolute 
magnitudes of the SC and AF gaps, as observed experimentally~\cite{za.ha.99}.

\section*{Acknowledgments}

This paper is dedicated to Professor Franz Wegner on the occasion of his
60$^{th}$ birthday.

We acknowledge many enlightening and pleasant discussions with S. C. Zhang.
This work was partially supported by the DFG (HA 1537/17-1).

\appendix

\section{Exact slave-boson treatment of the constraint}
\label{slav}

The hard-core constraint \eqref{hardcore} becomes (after projecting
out the electron pairs)
\beq
\label{cq}
Q(x)=  t_{\al}^{\dag}(x) t_{\al}(x) +
t_{h}^{\dag}(x) t_{h}(x)
+ e^{\dag}(x) e(x) - 1  =0 \;.
\eeq
The ``physical'' bosonic operators are then obtained as usual by the
replacement
\beq
\label{sbrep}
t_a(x) \to  t_a(x) e^{\dag}(x) \;,
\eeq
(including $a=h$)
so that the constraint is now conserved by the Hamiltonian.
Within the functional integral, the constraint \eqref{cq} can be
enforced as usual by adding a ``Lagrange multiplier'' term 
$i\sum_x \lambda(x) Q(x) $
and integrating over all $\lambda(x)$.
The partition function can thus be written in terms of an integral
over bosonic fields
\beq
\label{z}
{\cal Z} = \int \D t_{\al}^{\dag} \ \D t_{\al} \ \D t_h^{\dag} \ \D t_h \ \D e^{\dag} \ \D e \ d
\lam  \exp -S' \;,
\eeq
with the action 
\beqn
&&
S'=\int_{0}^{\beta} d \tau \ \Bigl\{ \sum_x  \Bigl[
t_{\al}^{\dag}(x,\tau) \left(\frac{\de}{\de \tau}+i \lambda(x)\right)
t_{\al}(x,\tau) +
t_h^{\dag}(x,\tau) \left(\frac{\de}{\de \tau}+i \lambda(x)\right)
t_h(x,\tau) 
\\ \nonumber &&
+
e^{\dag}(x,\tau) \left(\frac{\de}{\de \tau} +i\lambda(x)\right) e(x,\tau) 
-i \lambda(x) \Bigr] + H(\tau) \Bigr\} \;,
\eeqn
where $H(\tau)$ is obtained  by replacing \eqref{sbrep} in \eqref{ham}
and by replacing all bosonic operators 
with the corresponding fields
at the imaginary time $\tau$
(since the Hamiltonian is already normal ordered).
In principle, one should take a discretization of the time variable
and consider the continuum limit only at the end of the
calculation~\cite{be.prl,fi.phr,cc.prb,discr}. 
Notice that the integration of $\lambda$ would not give a constraint 
like \eqref{cq} for the bosonic fields at all imaginary times.
Nevertheless, one can proceed in the usual way by carrying out the gauge
transformation
\beqn
&& e(x,\tau) = \bar e(x,\tau) e^{i \theta(x,\tau)} 
\\ \nonumber &&
t_a(x,\tau) = \bar t_a(x,\tau) e^{i \theta(x,\tau)} 
\\ \nonumber &&
\lam(x) = \bar \lam(x,\tau) - \dot{\theta}(x,\tau) \;,
\eeqn
where
$\bar e(x,\tau) = | e(x,\tau) |$.
In this way, we can restrict to real values of the boson filed $e$ and 
absorb the time dependence of its phase into a (now) time-dependent
$\lam$.
Integration over $\lam(x,\tau)$ now leads to the enforcement of the
constraint via the $\delta$ function
(for simplicity, we drop the bar everywhere)
\beq
\prod_{x,\tau} \delta  \left[ 
|t_{\al}(x,\tau)|^2 + |t_h(x,\tau)|^2
 + e(x,\tau)^2 -1 \right]
\eeq
at all imaginary times. Integration over  $e(x,\tau)$
allows one to replace it everywhere in the Hamiltonian, leading to the new 
action \eqref{sqmscl} with \eqref{scl}.

\section{Integration of the momenta}
\label{mome}

The $\v p$-dependent part of the action has the general form
(Cf. \eqref{sqm} and \eqref{spm})
\beq
S_p = \int_0^{\be} d \tau \ \int d x\ \Delta \ A(|\v m(x)|^2) \ p_{\al}(x)^2 - i
p_{\al}(x) \ B(x) \;,
\eeq
where $A$ is a function of the superspin's magnitude squared (for simplicity, 
 we neglect gradient terms).
In order to absorb the coefficient $A$,
we define new momentum variables 
\beq
 p_{\al}'(x) = p_{\al}(x)  \sqrt{A(|\v m(x)|^2)} \;.
\eeq
However, since we don't want to produce an $m$-dependent Jacobian, we
carry out a similar transformation for the $m$ variables as
\beq
\label{mprime}
 m_a'(x) = m_a(x) g[|\v m(x)|^2] \;,
\eeq
where $g$ is chosen
in order to have a Jacobian equal to $1$.
This requirement gives the differential equation
\beq
A(|\v m|^2)^{3/2}  \left[ 
   g(|\v m|^2)^{n} - 2  |\v m|^2  g(|\v m|^2)^{n-1}  g'(|\v m|^2) \right] =1
 \;,
\eeq
$n$ ($=5$) being the number of components of the superspin $\v m$.
The solution of this  equation is 
\beq
\label{solg}
(\sqrt{r} g(r) )^n = \frac{n}{2} \int r^{n/2-1} A(r)^{-3/2} \ dr \;,
\eeq
where $r=|\v m|^2$.
Upon restricting to the lowest order of \eqref{cala},
$A(r) = 1 + \frac{J}{2 \Delta_s} r + {\cal O}(r^2)$,
 we obtain
\beq
g[|\v m(x)|^2] \approx ( 1- \frac{3 J }{7 \nDelta_s} |\v m(x)|^2 ) \;,
\eeq 
and its inverse \eqref{mamap}.

\def\nonformale#1{#1}
\def\formale#1{}
\def\spa{} \def\spb{}
\spa


\begin{thebibliography}{10}
\bibitem{burg}{
A similar action, obtained from an effective \sof symmetric fermionic model,
  has been obtained in Ref.~\onlinecite{bu.cl.97}. However, the Hubbard gap is
  not taken into account there.}%
\bibitem{discr}{
One should be aware of the fact that it is not completely correct to take the
  continuum limit in the functional integral at the outset (see
  Refs.~\cite{be.prl,fi.phr,cc.prb}. Different procedures for taking this limit
  may lead to a different value of the constraint (the ``1'' in \eqref{cond}),
  which is, thus, not well defined in the continuum limit. However, for
  low-energy properties we are interested in, the precise value of the
  constraint is not important.}%
\bibitem{norm1}{
Due to \eqref{e}, the magnitude of the superspin, rather than being constant,
  must be smaller than a constant.}%
\bibitem{aniso}{
The effect of the $c$-axis anisotropy of the high-\tc superconductors can be
  taken into account by a scale transformation in the $z$ ($c$) direction, i.
  e. by transforming to the variable $z'=\chi z$ with $\chi^2 = \frac{J_{xy} \
  a^2}{J_z \ c^2} >1$, where $J_{xy}$ ($J_z$) is the value of $J$ in the
  $ab$-plane ($c$-plane) (we assume that the ratio $\chi$ is the same for the
  AF and for the SC order parameter), and $c$ is the lattice spacing in the $z$
  direction. The effect of the transformation can then be reabsorbed by
  changing $\vv \to \vv \chi$ in the expressions \eqref{sgl}. This
  transformation implies that the correlation length in the $z$ direction will
  be reduced by a factor $\chi^{-1}$ with respect to the one in the $xy$
  direction.}%
\end{thebibliography}
\end{document}